\documentclass[aps,preprint]{revtex4}%
\usepackage{amsfonts}
\usepackage{amsmath}
\usepackage{amssymb}
\usepackage{graphicx}%
\begin{document}
\preprint{CTP-SCU/2015003}
\title{Minimal Length Effects on Tunnelling from Spherically Symmetric Black Holes}
\author{Benrong Mu$^{a,b}$}
\email{mubenrong@uestc.edu.cn}
\author{Peng Wang$^{b}$}
\email{pengw@scu.edu.cn}
\author{Haitang Yang$^{b,c}$}
\email{hyanga@scu.edu.cn}
\affiliation{$^{a}$School of Physical Electronics, University of Electronic Science and
Technology of China ,Chengdu, 610054, China}
\affiliation{$^{b}$Center for Theoretical Physics, College of Physical Science and
Technology, Sichuan University, Chengdu, 610064, PR China}
\affiliation{$^{c}$Kavli Institute for Theoretical Physics China (KITPC), Chinese Academy
of Sciences, Beijing 100080, P.R. China}

\begin{abstract}
In this paper, we investigate effects of the minimal length on quantum
tunnelling from spherically symmetric black holes using the Hamilton-Jacobi
method incorporating the minimal length. We first derive the deformed
Hamilton-Jacobi equations for scalars and fermions, both of which have the
same expressions. The minimal length correction to the Hawking temperature is
found to depend on the black hole's mass and the mass and angular momentum of
emitted particles. Finally, we calculate a Schwarzschild black hole's
luminosity and find the black hole evaporates to zero mass in infinite time.

\end{abstract}
\maketitle
\tableofcontents

%\affiliation{Center for Theoretical Physics, College of Physical Science and Technology,
%Sichuan University, Chengdu, 610064, PR China}

%\affiliation{Center for Theoretical Physics, College of Physical Science and Technology,
%Sichuan University, Chengdu, 610064, PR China}

\section{Introduction}

The classical theory of black holes predicts that anything, including light,
couldn't escape from the black holes. However, Stephen Hawking first showed
that quantum effects could allow black holes to emit particles. The formula of
Hawking temperature was first given in the frame of quantum field
theory\cite{Hawking:1974sw}. After that, various methods for deriving Hawking
radiation have been proposed. Among them is a semiclassical method of modeling
Hawking radiation as a tunneling effect proposed by Kraus and
Wilczek\cite{Kraus:1994by,Kraus:1994fj}, which is known as the null geodesic method. Later,
the tunneling behaviors of particles were investigated using the
Hamilton-Jacobi method\cite{Srinivasan:1998ty,Angheben:2005rm,Kerner:2006vu}. Using the null geodesic method
and the Hamilton-Jacobi method, much fruit has been achieved\cite{Hemming:2001we,Medved:2002zj,Vagenas:2001rm,Arzano:2005rs,Wu:2006pz,Nadalini:2005xp,Chatterjee:2007hc,
Akhmedova:2008dz,Akhmedov:2008ru,Akhmedova:2008au,Banerjee:2008ry,Singleton:2010gz}. The
key point of the Hamilton-Jacobi method is using WKB approximation to
calculate the imaginary part of the action for the tunneling process.

On the other hand, various theories of quantum gravity, such as string theory,
loop quantum gravity and quantum geometry, predict the existence of a minimal
length\cite{Townsend:1977xw,Amati:1988tn,Konishi:1989wk}. The generalized uncertainty
principle(GUP)\cite{Kempf:1994su} is a simply way to realize this minimal length.

\bigskip

An effective model of the GUP in one dimensional quantum mechanics is given
by\cite{Hossenfelder:2003jz,Hassan:2002qk}
\begin{equation}
L_{f}k(p)=\tanh\left(  \frac{p}{M_{f}}\right)  , \label{tanh1}%
\end{equation}%
\begin{equation}
L_{f}\omega(E)=\tanh\left(  \frac{E}{M_{f}}\right)  , \label{tanh2}%
\end{equation}
where the generators of the translations in space and time are the wave vector
$k$ and the frequency $\omega$, $L_{f}$ is the minimal length, and $L_{f}%
M_{f}=\hbar$. The quantization in position representation $\hat{x}=x$ leads
to
\begin{equation}
k=-i\partial_{x},\text{ }\omega=i\partial_{t}.
\end{equation}
Therefore, the low energy limit $p\ll M_{f}$ including order of $\frac{p^{3}%
}{M_{f}^{3}}$ gives
\begin{align}
p  &  =-i\hbar\partial_{x}\left(  1-\frac{\hbar^{2}}{M_{f}^{2}}\partial
_{x}^{2}\right)  ,\label{eq:momentum}\\
E  &  =i\hbar\partial_{t}\left(  1-\frac{\hbar^{2}}{M_{f}^{2}}\partial_{t}%
^{2}\right)  , \label{eq:energy}%
\end{align}
where we neglect the factor $\frac{1}{3}$. From eqn. $\left(  \ref{tanh1}%
\right)  $, it is noted that although one can increase $p$ arbitrarily, $k$
has an upper bound which is $\frac{1}{L_{f}}$. The upper bound on $k$ implies
that that particles could not possess arbitrarily small Compton wavelengths
$\lambda=2\pi/k$ and that there exists a minimal length $\sim L_{f}$.
Furthermore, the deformed Klein-Gordon/Dirac equations incorporating eqn.
$\left(  \ref{eq:momentum}\right)  $ and eqn. $\left(  \ref{eq:energy}\right)
$ have already be obtained in \cite{Hossenfelder:2003jz}, which will be
briefly reviewed in section \ref{Sec:DHJ}. So \cite{Hossenfelder:2003jz}
provides a way to incorporate the minimal length with special relativity, a
good starting point for studying Hawking radiation as tunnelling effect.

The black hole is a suitable venue to discuss the effects of quantum gravity.
Incorporating GUP into black holes has been discussed in a lot of
papers\cite{Ali:2012mt,Majumder:2011bv,Bina:2010ir,Chen:2002tu,Xiang:2009yq,Kim:2007hf}. The thermodynamics of black
holes has also been investigated in the framework of GUP\cite{Xiang:2009yq,Kim:2007hf}.
In \cite{Majhi:2013koa}, a new form of GUP is introduced%
\begin{align}
p^{0}  &  =k^{0},\\
p^{i}  &  =k^{i}\left(  1-\alpha\mathbf{k}+2\alpha^{2}\mathbf{k}^{2}\right)  ,
\end{align}
where $p^{a}$ is the modified four momentum, $k^{a}$ is the usual four
momentum and $\alpha$ is a small parameter. The modified velocity of photons,
two-dimensional Klein-Gordon equation the emission spectrum due to the Unruh
effect is obtained there. Recently, the GUP deformed Hamilton-Jacobi equation
for fermions in curved spacetime have been introduced and the corrected
Hawking temperatures have been derived\cite{Chen:2013pra,Chen:2013tha,Chen:2013ssa,Chen:2014xsa,liu:2014zxa}. The authors
consider the GUP of form%
\begin{align}
x_{i}  &  =x_{0i,}\\
p_{i}  &  =p_{0i}\left(  1+\beta p^{2}\right)  ,
\end{align}
where $x_{0i,}$ and $p_{0i}$ satisfy the canonical commutation relations.
Fermions' tunnelling, black hole thermodynamics and the remnants are discussed there.

In this paper, we investigate scalars and fermions tunneling across the
horizons of black holes using the deformed Hamilton-Jacobi method which
incorporates the minimal length via eqn. $\left(  \ref{eq:momentum}\right)  $
and eqn. $\left(  \ref{eq:energy}\right)  $. Our calculation shows that the
quantum gravity correction is related not only to the black hole's mass but
also to the mass and angular momentum of emitted particles.

The organization of this paper is as follows. In section \ref{Sec:DHJ}, from
the modified fundamental commutation relation, we generalize the
Hamilton-Jacobi in curved spacetime. In section \ref{Sec:PT}, incorporating
GUP, we investigate the tunnelling of particles in the black holes. In section
\ref{Sec:TBH}, we investigate how a Schwarzschild black hole evaporates in our
model. Section \ref{Sec:Con} is devoted to our conclusion. In this paper, we
take Geometrized units $c=G=1$, where the Planck constant $\hbar$ is square of
the Planck Mass $m_{p}$. We also assume that the emitted particles are neutral.

\section{Deformed Hamilton-Jacobi Equations}

\label{Sec:DHJ}

To be generic, we will consider a spherically symmetric background metric of
the form
\begin{equation}
ds^{2}=f\left(  r\right)  dt^{2}-\frac{dr^{2}}{f\left(  r\right)  }%
-r^{2}\left(  d\theta^{2}+\sin^{2}\theta d\phi^{2}\right)  ,
\label{Schwarzschild metric}%
\end{equation}
where where $f\left(  r\right)  $ has a simple zero at $r=r_{h}$ with
$f^{\prime}\left(  r_{h}\right)  $ being finite and nonzero. The vanishing of
$f\left(  r\right)  $ at point $r=r_{h}$ indicates the presence of an event
horizon. In this section, we will first derive the deformed Klein-Gordon/Dirac
equations in flat spacetime and then generalize them to the curved spacetime
with the metric $\left(  \ref{Schwarzschild metric}\right)  $.

In the $\left(  3+1\right)  $ dimensional flat spacetime, the relations
between $\left(  p_{i},E\right)  $ and $\left(  k_{i},\omega\right)  $ can
simply be generalized to%
\begin{align}
L_{f}k_{i}(p)  &  =\tanh\left(  \frac{p_{i}}{M_{f}}\right)
,\label{eq:wavevector}\\
L_{f}\omega(E)  &  =\tanh\left(  \frac{E}{M_{f}}\right)  ,
\label{eq:frequency}%
\end{align}
where, in the spherical coordinates, one has for $\vec{k}$%
\begin{equation}
\vec{k}=-i\left(  \hat{r}\frac{\partial}{\partial r}+\frac{\hat{\theta}}%
{r}\frac{\partial}{\partial\theta}+\frac{\hat{\phi}}{r\sin\theta}%
\frac{\partial}{\partial\phi}\right)  .
\end{equation}
Expanding eqn. $\left(  \ref{eq:wavevector}\right)  $ and eqn. $\left(
\ref{eq:frequency}\right)  $ for small arguments to the third order gives the
energy and momentum operator in position representation%
\begin{gather}
E=i\hbar\partial_{t}\left(  1-\frac{\hbar^{2}}{M_{f}^{2}}\partial_{t}%
^{2}\right)  ,\\
\vec{p}=\frac{\hbar}{i}\left[  \hat{r}\left(  \partial_{r}-\frac{\hbar
^{2}\partial_{r}^{3}}{M_{f}^{2}}\right)  +\hat{\theta}\left(  \frac
{\partial_{\theta}}{r}-\frac{\hbar^{2}\partial_{\theta}^{3}}{r^{3}M_{f}^{2}%
}\right)  +\hat{\phi}\left(  \frac{\partial_{\phi}}{r\sin\theta}-\frac
{\hbar^{2}\partial_{\phi}^{3}}{r^{3}\sin^{3}\theta M_{f}^{2}}\right)  \right]
,
\end{gather}
where we also omit the factor $\frac{1}{3}$. Substituting the above energy and
momentum operators into the energy-momentum relation, the deformed
Klein-Gordon equation satisfied by the scalar field with the mass $m$ is
\begin{equation}
E^{2}\phi=p^{2}\phi+m^{2}\phi, \label{eq:KGFD}%
\end{equation}
where $p^{2}=\vec{p}\cdot\vec{p}.$ Making the ansatz for $\phi$
\begin{equation}
\phi=\exp\left(  \frac{iI}{\hbar}\right)  ,
\end{equation}
and substituting it into eqn. $\left(  \ref{eq:KGFD}\right)  $, one expands
eqn. $\left(  \ref{eq:KGFD}\right)  $ in powers of $\hbar$ and then finds that
the lowest order gives the deformed scalar Hamilton-Jacobi equation in the
flat spacetime
\begin{gather}
\left(  \partial_{t}I\right)  ^{2}\left(  1+\frac{2\left(  \partial
_{t}I\right)  ^{2}}{M_{f}^{2}}\right)  -\left(  \partial_{r}I\right)
^{2}\left(  1+\frac{2\left(  \partial_{r}I\right)  ^{2}}{M_{f}^{2}}\right)
-\frac{\left(  \partial_{\theta}I\right)  ^{2}}{r^{2}}\left(  1+\frac{2\left(
\partial_{\theta}I\right)  ^{2}}{r^{2}M_{f}^{2}}\right) \nonumber\\
-\frac{\left(  \partial_{\phi}I\right)  ^{2}}{r^{2}\sin^{2}\theta}\left(
1+\frac{2\left(  \partial_{\phi}I\right)  ^{2}}{r^{2}\sin^{2}\theta M_{f}^{2}%
}\right)  =m^{2}, \label{eq:HJFD}%
\end{gather}
which is truncated at $\mathcal{O}\left(  \frac{1}{M_{f}^{2}}\right)  $.

Similarly, the deformed Dirac equation for a spin-$1/2$ fermion with the mass
$m$ takes the form as%
\begin{equation}
\left(  \gamma_{0}E+\vec{\gamma}\cdot\vec{p}-m\right)  \psi=0,
\label{eq:DiracFD}%
\end{equation}
where $\left\{  \gamma_{0},\gamma_{0}\right\}  =2$, $\left\{  \gamma
_{a},\gamma_{b}\right\}  =-2\delta_{ab}$, and $\left\{  \gamma_{0},\gamma
_{a}\right\}  =0$ with the Latin index $a$ running over $r,\theta,$and $\phi$.
Multiplying $\left(  \gamma_{0}E+\vec{\gamma}\cdot\vec{p}+m\right)  $ by eqn.
$\left(  \ref{eq:DiracFD}\right)  $ and using the gamma matrices
anticommutation relations, the deformed Dirac equation can be written as%
\begin{equation}
E^{2}\psi=\left(  p^{2}+m^{2}\right)  \psi-\frac{\left[  \gamma_{a},\gamma
_{b}\right]  }{2}p_{a}p_{b}\psi. \label{eq:DiracFDM}%
\end{equation}
To obtain the Hamilton-Jacobi equation for the fermion, the ansatz for $\psi$
takes the form of
\begin{equation}
\psi=\exp\left(  \frac{iI}{\hbar}\right)  v, \label{eq:fermionansatzD}%
\end{equation}
where $v$ is a vector function of the spacetime. Substituting eqn. $\left(
\ref{eq:fermionansatzD}\right)  $ into eqn. $\left(  \ref{eq:DiracFDM}\right)
$ and noting that the second term on RHS of eqn. $\left(  \ref{eq:DiracFDM}%
\right)  $ does not contribute to the lowest order of $\hbar$, we find the
deformed Hamilton-Jacobi equation for a fermion is the same as the deformed
one for a scalar with the same mass, namely eqn. $\left(  \ref{eq:HJFD}%
\right)  $. Note that one can use the deformed Maxwell's equations obtained in
\cite{Hossenfelder:2003jz} to get the deformed Hamilton-Jacobi equation for a
vector boson. However, for simplicity we just stop here.

In order to generalize the deformed Hamilton-Jacobi equation, eqn. $\left(
\ref{eq:HJFD}\right)  $, to the curved spacetime with the metric $\left(
\ref{Schwarzschild metric}\right)  $, we first consider the Hamilton-Jacobi
equation without GUP modifications. In ref. \cite{Benrong:2014woa}, we show
that the unmodified Hamilton-Jacobi equation in curved spacetime with
$ds^{2}=g_{\mu\nu}dx^{\mu}dx^{\nu}$ is%
\begin{equation}
g^{\mu\nu}\partial_{\mu}I\partial_{\nu}I-m^{2}=0.
\end{equation}
Therefore, the unmodified Hamilton-Jacobi equation in the metric $\left(
\ref{Schwarzschild metric}\right)  $ becomes
\begin{equation}
\frac{\left(  \partial_{t}I\right)  ^{2}}{f\left(  r\right)  }-f\left(
r\right)  \left(  \partial_{r}I\right)  ^{2}-\frac{\left(  \partial_{\theta
}I\right)  ^{2}}{r^{2}}-\frac{\left(  \partial_{\phi}I\right)  ^{2}}{r^{2}%
\sin^{2}\theta}=m^{2}. \label{eq:HJC}%
\end{equation}
On the other hand, the unmodified Hamilton-Jacobi equation in flat spacetime
can be obtained from eqn. $\left(  \ref{eq:HJFD}\right)  $ by taking
$M_{f}\rightarrow\infty$,%
\begin{equation}
\left(  \partial_{t}I\right)  ^{2}-\left(  \partial_{r}I\right)  ^{2}%
-\frac{\left(  \partial_{\theta}I\right)  ^{2}}{r^{2}}-\frac{\left(
\partial_{\phi}I\right)  ^{2}}{r^{2}\sin^{2}\theta}=m^{2}. \label{eq:HJFF}%
\end{equation}
Comparing eqn. $\left(  \ref{eq:HJC}\right)  $ with eqn. $\left(
\ref{eq:HJFF}\right)  $, one finds that the Hamilton-Jacobi equation in the
metric $\left(  \ref{Schwarzschild metric}\right)  $ can be obtained from the
one in flat spacetime by making replacements $\partial_{r}I\rightarrow
\sqrt{f\left(  r\right)  }\partial_{r}I$ and $\partial_{t}I\rightarrow
\frac{\partial_{t}I}{\sqrt{f\left(  r\right)  }}$ in the no GUP modifications
scenario. Therefore, by making replacements $\partial_{r}I\rightarrow
\sqrt{f\left(  r\right)  }\partial_{r}I$ and $\partial_{t}I\rightarrow
\frac{\partial_{t}I}{\sqrt{f\left(  r\right)  }}$, the deformed
Hamilton-Jacobi equation in flat spacetime, eqn. $\left(  \ref{eq:HJFD}%
\right)  $, leads to the deformed Hamilton-Jacobi equation in the metric
$\left(  \ref{Schwarzschild metric}\right)  $, which is to $\mathcal{O}\left(
\frac{1}{M_{f}^{2}}\right)  ,$
\begin{gather}
\frac{\left(  \partial_{t}I\right)  ^{2}}{f\left(  r\right)  }\left(
1+\frac{2\left(  \partial_{t}I\right)  ^{2}}{f\left(  r\right)  M_{f}^{2}%
}\right)  -f\left(  r\right)  \left(  \partial_{r}I\right)  ^{2}\left(
1+\frac{2f\left(  r\right)  \left(  \partial_{r}I\right)  ^{2}}{M_{f}^{2}%
}\right)  -\frac{\left(  \partial_{\theta}I\right)  ^{2}}{r^{2}}\left(
1+\frac{2\left(  \partial_{\theta}I\right)  ^{2}}{r^{2}M_{f}^{2}}\right)
\nonumber\\
-\frac{\left(  \partial_{\phi}I\right)  ^{2}}{r^{2}\sin^{2}\theta}\left(
1+\frac{2\left(  \partial_{\phi}I\right)  ^{2}}{r^{2}\sin^{2}\theta M_{f}^{2}%
}\right)  =m^{2}. \label{eq:HJCG}%
\end{gather}

\section{Quantum Tunnelling}

\label{Sec:PT}

In this section, we investigate the particles' tunneling at the event horizon
$r=r_{h}$ of the metric $\left(  \ref{Schwarzschild metric}\right)  $ where
GUP is taken into account. Since the metric $\left(
\ref{Schwarzschild metric}\right)  $ does not depend on $t$ and $\phi$,
$\partial_{t}$ and $\partial_{\phi}$ are killing vectors. Taking into account
the Killing vectors of the background spacetime, we can employ the following
ansatz for the action
\begin{equation}
I=-\omega t+W\left(  r,\theta\right)  +p_{\phi}\phi, \label{eq:Ianzatz}%
\end{equation}
where $\omega$ and $p_{\phi}$ are constants and they are the energy and the
$z$-component of angular momentum of emitted particles, respectively.
Inserting eqn. $(\ref{eq:Ianzatz})$ into eqn. $(\ref{eq:HJCG}),$ we find that
the deformed Hamilton-Jacobi equation becomes
\begin{equation}
p_{r}^{2}\left(  1+\frac{2f\left(  r\right)  p_{r}^{2}}{M_{f}^{2}}\right)
=\frac{1}{f^{2}\left(  r\right)  }\left[  \omega^{2}\left(  1+\frac
{2\omega^{2}}{f\left(  r\right)  M_{f}^{2}}\right)  -f\left(  r\right)
\left(  m^{2}+\lambda\right)  \right]  , \label{eq:pr}%
\end{equation}
where $p_{r}=\partial_{r}W$, $p_{\theta}=\partial_{\theta}W$ and
\[
\lambda=\frac{p_{\theta}^{2}}{r^{2}}\left(  1+\frac{2p_{\theta}^{2}}%
{r^{2}M_{f}^{2}}\right)  +\frac{p_{\phi}^{2}}{r^{2}\sin^{2}\theta}\left(
1+\frac{2p_{\phi}^{2}}{r^{2}\sin^{2}\theta M_{f}^{2}}\right)  .
\]
Since the magnitude of the angular momentum of the particle $L$ can be
expressed in terms of $p_{\theta}$ and $p_{\phi},$
\begin{equation}
L^{2}=p_{\theta}^{2}+\frac{p_{\phi}^{2}}{\sin^{2}\theta},
\end{equation}
one can rewrite $\lambda$ as%
\begin{equation}
\lambda=\frac{L^{2}}{r^{2}}+\mathcal{O}\left(  \frac{1}{M_{f}^{2}}\right)  .
\label{eq:lamda}%
\end{equation}
Solving eqn. $(\ref{eq:pr})$ for $p_{r}$ to $\mathcal{O}\left(  \frac{1}%
{M_{f}^{2}}\right)  $gives
\begin{align}
\partial_{r}W_{\mp}  &  =\pm\frac{1}{f\left(  r\right)  }\sqrt{\omega
^{2}\left(  1+\frac{2\omega^{2}}{f\left(  r\right)  M_{f}^{2}}\right)
-f\left(  r\right)  \left(  m^{2}+\lambda\right)  }\times\nonumber\\
&  \sqrt{1-\frac{2}{M_{f}^{2}f\left(  r\right)  }\left[  \omega^{2}\left(
1+\frac{2\omega^{2}}{f\left(  r\right)  M_{f}^{2}}\right)  -f\left(  r\right)
\left(  m^{2}+\lambda\right)  \right]  }, \label{eq:dW}%
\end{align}
where +/$-$ represent the outgoing/ingoing solutions. In order to get the
imaginary part of $W_{\pm}$, we need to find residue of the RHS of eqn.
$(\ref{eq:dW})$ at $r=r_{h}$ by expanding the RHS in a Laurent series with
respect to $r$ at $r=r_{h}$. We then rewrite eqn. $(\ref{eq:dW})$ as
\begin{align}
\partial_{r}W_{\mp}  &  =\pm\frac{1}{f\left(  r\right)  ^{\frac{5}{2}}}%
\sqrt{\omega^{2}\left(  f\left(  r\right)  +\frac{2\omega^{2}}{M_{f}^{2}%
}\right)  -f^{2}\left(  r\right)  \left(  m^{2}+\lambda\right)  }%
\times\nonumber\\
&  \sqrt{f^{2}\left(  r\right)  -\frac{2}{M_{f}^{2}}\left[  \omega^{2}\left(
f\left(  r\right)  +\frac{2\omega^{2}}{M_{f}^{2}}\right)  -f^{2}\left(
r\right)  \left(  m^{2}+\lambda\right)  \right]  }.
\end{align}
Using $f\left(  r\right)  =f^{\prime}\left(  r_{h}\right)  \left(
r-r_{h}\right)  +\frac{f^{\prime\prime}\left(  r_{h}\right)  }{2}\left(
r-r_{h}\right)  ^{2}+\mathcal{O}\left(  \left(  r-r_{h}\right)  ^{3}\right)
$, one can single out the $\frac{1}{r-r_{h}}$ term of the Laurent series
\begin{equation}
\partial_{r}W_{\mp}\sim\frac{\pm a_{-1}}{r-r_{h}},
\end{equation}
where we have
\begin{equation}
a_{-1}=\frac{\omega}{f^{\prime}\left(  r_{h}\right)  }\left[  1+\frac{2}%
{M_{f}^{2}}\left(  m^{2}+\frac{L^{2}}{r_{h}^{2}}\right)  \right]
+\mathcal{O}\left(  \frac{1}{M_{f}^{4}}\right)  .
\end{equation}
Using the residue theory for semi circles, we obtain for the imaginary part of
$W_{\pm}$ to $\mathcal{O}\left(  \frac{1}{M_{f}^{2}}\right)  $
\begin{equation}
\operatorname{Im}W_{\pm}=\pm\frac{\pi\omega}{f^{\prime}\left(  r_{h}\right)
}\left[  1+\frac{2}{M_{f}^{2}}\left(  m^{2}+\frac{L^{2}}{r_{h}^{2}}\right)
\right]  .
\end{equation}

However, when one tries to calculate the tunneling rate $\Gamma$ from
$\operatorname{Im}W_{\pm}$, there is so called \textquotedblleft factor-two
problem\textquotedblright\cite{Akhmedov:2006pg}. One way to solve the "factor
two problem" is introducing a \textquotedblleft temporal
contribution\textquotedblright%
\cite{Akhmedova:2008au,Akhmedova:2008dz,Akhmedov:2008ru,Chowdhury:2006sk,Akhmedov:2006un}%
. To consider an invariance under canonical transformations, we also follow
the recent
work\cite{Akhmedova:2008au,Akhmedova:2008dz,Akhmedov:2008ru,Chowdhury:2006sk,Akhmedov:2006un}
and adopt the expression $%
%TCIMACRO{\doint }%
%BeginExpansion
{\displaystyle\oint}
%EndExpansion
p_{r}dr=\int p_{r}^{+}dr-\int p_{r}^{-}dr$ for the spatial contribution to
$\Gamma$. The spatial and temporal contributions to $\Gamma$ are given as follows.

\textbf{Spatial Contribution:} The spatial part contribution comes from the
imaginary part of $W\left(  r\right)  $. Thus, the spatial part contribution
is proportional to
\begin{align}
&  exp\left[  -\frac{1}{\hbar}\operatorname{Im}\oint p_{r}dr\right]
\nonumber\\
&  =exp\left[  -\frac{1}{\hbar}\operatorname{Im}\left(  \int p_{r}^{+}dr-\int
p_{r}^{-}dr\right)  \right] \nonumber\\
&  =exp\left\{  -\frac{2\pi\omega}{f^{\prime}\left(  r_{h}\right)  }\left[
1+\frac{2}{M_{f}^{2}}\left(  m^{2}+\frac{L^{2}}{r_{h}^{2}}\right)  \right]
\right\}
\end{align}

\textbf{Temporal Contribution:} As pointed in Ref.
\cite{Akhmedov:2008ru,Akhmedov:2006pg,Chowdhury:2006sk,Akhmedov:2006un}, the
temporal part contribution comes from the "rotation" which connects the
interior region and the exterior region of the black hole. Thus, the imaginary
contribution from the temporal part when crossing the horizon is
$\operatorname{Im}\left(  \omega\Delta t^{out,in}\right)  =\omega\frac{\pi
}{2\kappa}$, where $\kappa=\frac{f^{\prime}\left(  r_{h}\right)  }{2}$ is the
surface gravity at the event horizon. Then the total temporal contribution for
a round trip is
\begin{equation}
\operatorname{Im}\left(  \omega\Delta t\right)  =\frac{2\pi\omega}{f^{\prime
}\left(  r_{h}\right)  }.
\end{equation}
Therefore, the tunnelling rate of the particle crossing the horizon is
\begin{align}
\Gamma &  \propto exp\left[  -\frac{1}{\hbar}\left(  \operatorname{Im}\left(
\omega\Delta t\right)  +\operatorname{Im}\oint p_{r}dr\right)  \right]
\nonumber\\
&  =exp\left\{  -\frac{4\pi\omega}{f^{\prime}\left(  r_{h}\right)  }\left[
1+\frac{1}{M_{f}^{2}}\left(  m^{2}+\frac{L^{2}}{r_{h}^{2}}\right)  \right]
\right\}  . \label{eq:Gamma}%
\end{align}
This is the expression of Boltzmann factor with an effective temperature
\begin{equation}
T=\frac{f^{\prime}\left(  r_{h}\right)  }{4\pi}\frac{\hbar}{1+\frac{1}%
{M_{f}^{2}}\left(  m^{2}+\frac{L^{2}}{r_{h}^{2}}\right)  }, \label{HKT1}%
\end{equation}
where $T_{0}=\frac{\hbar f^{\prime}\left(  r_{h}\right)  }{4\pi}$ is the
original Hawking temperature.

For the standard Hawing radiation, all particles very close to the horizon are
effectively massless on account of infinite blueshift. Thus, the conformal
invariance of the horizon make Hawing temperatures of all particles the same.
The mass, angular momentum and identity of the particles are only relevant
when they escape the potential barrier. However, if quantum gravity effects
are considered, behaviors of particles near the horizon could be different.
For example, if we send a wave packet which is governed by a subluminal
dispersion relation, backwards in time toward the horizon, it reaches a
minimum distance of approach, then reverse direction and propagate back away
from the horizon, instead of getting unlimited blueshift toward the horizon
\cite{Unruh:1994je,Corley:1996ar}. Thus, quantum gravity effects might make fermions and
scalars experience different (effective) Hawking temperatures. However, our
result shows that in our model, the tunnelling rates of fermions and scalars
depend on their masses and angular momentums, but independent of the
identities of the particles, to $\mathcal{O}\left(  \frac{1}{M_{f}^{2}%
}\right)  $. In other words, effective Hawking temperatures of fermions and
scalars are the same to $\mathcal{O}\left(  \frac{1}{M_{f}^{2}}\right)  $ in
our model as long as their masses, energies and angular momentums are the same.

\section{Thermodynamics of Black Holes}

\label{Sec:TBH}

For simplicity, we consider the Schwarzschild metric with $f\left(  r\right)
=1-\frac{2M}{r}$ with the black hole's mass, $M$. The event horizon of the
Schwarzschild black hole is $r_{h}=2M$. In this section, we work with massless
particles. Near the horizon of the the black hole, angular momentum of the
particle $L\sim pr_{h}\sim\omega r_{h}$. Thus, one can rewrite $T$%
\begin{equation}
T\sim\frac{T_{0}}{1+\frac{2\omega^{2}}{M_{f}^{2}}}, \label{eq:Temp}%
\end{equation}
where $T_{0}=\frac{\hbar}{8\pi M}$ for the Schwarzschild black hole. As
reported in \cite{AmelinoCamelia:2005ik}, the authors obtained the
relation $\omega\gtrsim\frac{\hbar}{\delta x}$ between the energy of a
particle and its position uncertainty in the framework of GUP. Near the
horizon of the the Schwarzschild black hole, the position uncertainty of a
particle will be of the order of the Schwarzschild radius of the black hole
\cite{Bekenstein:1973ur} $\delta x\sim r_{h}$. Thus, one finds for $T$%
\begin{equation}
T\sim\frac{T_{0}}{1+\frac{m_{p}^{4}}{2M^{2}M_{f}^{2}}}, \label{eq: Temp}%
\end{equation}
where we use $\hbar=m_{p}^{2}$. Using the first law of the black hole
thermodynamics, we find the corrected black hole entropy is%
\begin{align}
S  &  =\int\frac{dM}{T}\nonumber\\
&  \sim\frac{A}{4m_{p}^{2}}+\frac{4\pi m_{p}^{2}}{M_{f}^{2}}\ln\left(
\frac{A}{16\pi}\right)  , \label{eq:entropy}%
\end{align}
where $A=4\pi r_{h}^{2}=16\pi M^{2}$ is the area of the horizon. The
logarithmic term in eqn. $\left(  \ref{eq:entropy}\right)  $ is the well known
correction from quantum gravity to the classical Bekenstein-Hawking entropy,
which have appeared in different studies of GUP modified thermodynamics of
black holes\cite{Bina:2010ir,Chen:2002tu,Xiang:2009yq,Majumder:2011xg,Nozari:2012nf,
Banerjee:2010sd,Cavaglia:2004jw,Myung:2006qr,Nouicer:2007jg,Majumder:afa}. In general,
the entropy for the Schwarzschild black hole of mass $M$ in four spacetime
dimensions can be written in form of%
\begin{equation}
S=\frac{A}{4}+\sigma\ln\left(  \frac{A}{16\pi}\right)  +\mathcal{O}\left(
\frac{M_{f}^{2}}{A}\right)  , \label{eq:EntropyExpansion}%
\end{equation}
where $\sigma=\frac{2M_{f}^{2}}{M^{2}}$ in our paper. Neglecting the terms
$\mathcal{O}\left(  \frac{M_{f}^{2}}{A}\right)  $ in eqn. $\left(
\ref{eq:EntropyExpansion}\right)  \,$, there could be three scenarios
depending on the sign of $\sigma$.

\begin{enumerate}
\item $\sigma=0:$ This case is just the standard Hawking radiation. The black
holes evaporate completely in finite time.

\item $\sigma<0:$ The entropy $S$ as function of mass develops a minimum at
some value of $M_{\min}$. This predicts the existence of black hole remnants.
Furthermore, the black holes stop evaporating in finite time. This is what
happened in \cite{Bina:2010ir,Chen:2002tu,Xiang:2009yq,Majumder:2011xg,Nozari:2012nf,
Banerjee:2010sd,Cavaglia:2004jw,Myung:2006qr,Nouicer:2007jg,Majumder:afa}, which is
consistent with the existence of a minimal length.

\item $\sigma>0:$ This is a subtle case. In the remaining of the section, we
will investigate how the black holes evaporates in our model.
\end{enumerate}

For particles emitted in a wave mode labelled by energy $\omega$ and $L,$ we
find from eqn. $\left(  \ref{eq:Gamma}\right)  $ that \cite{Hartle:1976tp}%
\begin{align*}
&  \left(  \text{Probability for a black hole to emit a particle in this
mode}\right) \\
&  =\exp\left(  -\frac{\omega}{T}\right)  \times(\text{Probability for a black
hole to absorb a particle in the same mode}),
\end{align*}
where $T$ is given by eqn. $\left(  \ref{HKT1}\right)  $. Neglecting
back-reaction, detailed balance condition requires that the ratio of the
probability of having $N$ particles in a particular mode with $\omega$ and $L$
to the probability of having $N-1$ particles in the same mode is $\exp\left(
-\frac{\omega}{T}\right)  .$ One then follows the standard textbook procedure
to get the average number $n_{\omega,L}$ in the mode%
\begin{equation}
n_{\omega,L}=n\left(  \frac{\omega}{T}\right)  ,
\end{equation}
where we define%
\begin{equation}
n\left(  x\right)  =\frac{1}{\exp x-\left(  -1\right)  ^{\epsilon}},
\end{equation}
and $\epsilon=0$ for bosons and $\epsilon=1$ for fermions. In
\cite{Page:1976df}, counting the number of modes per frequency interval
with periodic boundary conditions in a large container around the black hole,
Page related the expected number emitted per mode $n_{\omega,L}$ to the
average emission rate per frequency interval $\frac{dn_{\omega,L}}{dt}$ by%
\begin{equation}
\frac{dn_{\omega,L}}{dt}=n_{\omega,L}\frac{d\omega}{2\pi\hbar}.
\label{eq:DnDt}%
\end{equation}
Following the Page's argument, we find that in our model
\begin{equation}
\frac{dn_{\omega,L}}{dt}=n_{\omega,L}\frac{\partial\omega}{\partial p_{r}%
}\frac{dp_{r}}{2\pi\hbar}=n_{\omega,L}\frac{d\omega}{2\pi\hbar},
\label{eq:DnDt-MDR}%
\end{equation}
where $\frac{\partial\omega}{\partial p_{r}}$ is the radial velocity of the
particle and the number of modes between the wavevector interval $\left(
p_{r},p_{r}+dp_{r}\right)  $ is $\frac{dp_{r}}{2\pi\hbar}$ where
$p_{r}=\partial_{r}I$ is the radial wavevector. Since each particle carries
off the energy $\omega$, the total luminosity is obtained from $\frac
{dn_{\omega,L}}{dt}$ by multiplying by the energy $\omega$ and summing up over
all energy $\omega$ and $L$,
\begin{equation}
L=%
%TCIMACRO{\dsum \limits_{l=0}}%
%BeginExpansion
{\displaystyle\sum\limits_{l=0}}
%EndExpansion
\left(  2l+1\right)  \int\omega n_{\omega,l}\frac{d\omega}{2\pi\hbar},
\end{equation}
where $L^{2}=\left(  l+1\right)  l\hbar^{2}$ and the degeneracy for $l$ is
$\left(  2l+1\right)  $. However, some of the radiation emitted by the horizon
might not be able to reach the asymptotic region. We need to consider the
greybody factor $\left\vert T_{i}\left(  \omega\right)  \right\vert ^{2}$,
where $T_{i}\left(  \omega\right)  $ represents the transmission coefficient
of the black hole barrier which in general can depend on the energy $\omega$
and angular momentum $l$ of the particle. Taking the greybody factor into
account, we find for the total luminosity%
\begin{equation}
L=%
%TCIMACRO{\dsum \limits_{l=0}}%
%BeginExpansion
{\displaystyle\sum\limits_{l=0}}
%EndExpansion
\left(  2l+1\right)  \int\left\vert T_{i}\left(  \omega\right)  \right\vert
^{2}\omega n_{\omega,l}\frac{d\omega}{2\pi\hbar}.
\end{equation}
Usually, one needs to solve the exact wave equations for $\left\vert
T_{i}\left(  \omega\right)  \right\vert ^{2},$ which is very complicated. On
the other hand, one can use the geometric optics approximation to estimate
$\left\vert T_{i}\left(  \omega\right)  \right\vert ^{2}$. In the geometric
optics approximation, we assume $\omega\gg M$ and high energy waves will be
absorbed unless they are aimed away from the black hole. Hence we have
$\left\vert T_{i}\left(  \omega\right)  \right\vert ^{2}=1$ for all the
classically allowed energy $\omega$ and angular momentum $l$ of the particle.
In this approximation, the Schwarzschild black hole is just like a black
sphere of radius $R=3^{3/2}M$ \cite{Wald:1984rg}, which puts an upper
bound on $l\left(  l+1\right)  \hbar^{2},$%
\begin{equation}
l\left(  l+1\right)  \hbar^{2}\leqslant27M^{2}\omega^{2}.
\end{equation}
Note that we neglect possible modifications from GUP to the bound since we are
interested in the GUP effects near the horizon. Thus, the luminosity is%
\begin{align}
L  &  =\int_{0}^{\infty}\frac{\omega d\omega}{2\pi\hbar^{3}}\int_{0}%
^{27M^{2}\omega^{2}}n\left[  \frac{\omega}{T_{0}}\left(  1+\frac{1}{M_{f}^{2}%
}\frac{l\left(  l+1\right)  \hbar^{2}}{2M^{2}}\right)  \right]  d\left[
l\left(  l+1\right)  \hbar^{2}\right] \nonumber\\
&  =\frac{T_{0}^{4}M^{2}}{2\pi\hbar^{3}}\int_{0}^{\infty}u^{3}du\int_{0}%
^{27}n\left[  u\left(  1+a^{2}ux\right)  \right]  dx, \label{eq:integral}%
\end{align}
where we define $u=\frac{\omega}{T_{0}}$, $y=\frac{l\left(  l+1\right)
\hbar^{2}}{M^{2}\omega^{2}}$ and $a=\frac{T_{0}}{\sqrt{2}M_{f}}=\frac
{m_{p}^{2}}{8\sqrt{2}\pi M_{f}M}$. For $M\gg\frac{m_{p}^{2}}{8\sqrt{2}\pi
M_{f}}$, we have $a\ll1$ and hence the luminosity is%
\begin{equation}
L\approx\frac{27}{32\pi^{2}\hbar^{3}}T_{0}^{4}A\int_{0}^{\infty}u^{3}n\left(
u\right)  du,
\end{equation}
which is just the Stefan's law for black holes. Therefore for large black
holes, they evaporate in almost the same way as in Case 1 until $M\sim
\frac{m_{p}^{2}}{8\sqrt{2}\pi M_{f}}$, when the term $a^{2}ux$ starts to
dominate in eqn. $\left(  \ref{eq:integral}\right)  $. Then the luminosity is
approximated by
\begin{align}
L  &  \sim\frac{T_{0}^{4}M^{2}}{2\pi\hbar^{3}}\int_{0}^{\infty}u^{3}du\int
_{0}^{27}n\left(  a^{2}u^{2}x\right)  dx\nonumber\\
&  =\frac{2M^{2}M_{f}^{4}}{\pi m_{p}^{6}}\int_{0}^{\infty}v^{3}dv\int_{0}%
^{27}n\left(  v^{2}x\right)  dx,
\end{align}
where $v=au$. Not worrying about exact numerical factors, one has for the
evaporation rate
\begin{equation}
\frac{dM}{dt}=-L\sim-A\frac{M^{2}}{M_{f}^{2}}, \label{eq:EvaporationRate}%
\end{equation}
where $A>0$ is a constant. Solving eqn. $\left(  \ref{eq:EvaporationRate}%
\right)  $ for $M$ gives $M\sim\frac{M_{f}^{2}}{At}$. The evaporation rate
considerably slows down when black holes' mass $M\sim\frac{m_{p}^{2}}%
{8\sqrt{2}\pi M_{f}}$. The black hole then evaporates to zero mass in infinite
time. However, the GUP predicts the existence of a minimal length. It would
make much more sense if there are black hole remnants in the GUP models. How
can we reconcile the contradiction? When we write down the deformed
Hamilton-Jacobi equation, eqn. $\left(  \ref{eq:HJCG}\right)  $, we neglect
terms higher than $\mathcal{O}\left(  \frac{1}{M_{f}^{2}}\right)  $. However,
when $M\sim\frac{m_{p}^{2}}{8\sqrt{2}\pi M_{f}}$, our effective approach
starts breaking down since contributions from higher order terms become the
same important as these from terms $\mathcal{O}\left(  \frac{1}{M_{f}^{2}%
}\right)  $. Thus, one then has to include these higher order
contributions.\ For example, if there are higher order corrections to eqn.
$\left(  \ref{eq:EvaporationRate}\right)  $ as in
\begin{equation}
\frac{dM}{dt}\sim-A\frac{M^{2}}{M_{f}^{2}}+B\frac{M^{3}}{M_{f}^{3}},
\end{equation}
where $B>0$, one can easily see that there exists a minimum mass $M_{\min}\sim
M_{f}\sqrt{\frac{A}{B}}$\ for black holes. In another word, the $\mathcal{O}%
\left(  \frac{1}{M_{f}^{2}}\right)  $ terms used in our paper are not
sufficient enough to produce black hole remnants predicted by the GUP. To do
so, one needs to resort to higher order terms if the full theory is not
available. It is also noted that when at late stage of a black hole with
$a\gtrsim1\,$, eqn. $\left(  \ref{eq:Temp}\right)  $ becomes
\begin{equation}
T\sim\frac{T_{0}}{\frac{m_{p}^{4}}{2M^{2}M_{f}^{2}}}\sim M,
\end{equation}
which means the tempreture of the black hole goes to zero as the mass goes to zero.

The GUP is closely related to noncommutative geometry. In fact, when the GUP
is investigated in more than one dimension, a noncommutative geometric
generalization of position space always appears naturally\cite{Kempf:1994su}.
On the other hand, quantum black hole physics has been studied in the
noncommutative geometry\cite{Nicolini:2008aj}. In
\cite{Banerjee:2008gc}, a noncommutative black hole's entropy received a
logarithmic correction with $\sigma<0$. However, \cite{Banerjee:2008du,Spallucci:2009zz}
showed that the corrections to a noncommutative schwarzschild black hole's
entropy might not involve any logarithmic terms. In either case, the
tempreture of the noncommutative schwarzschild black hole reaches zero in
finite time with remnants left.

\section{ Conclusion}

\label{Sec:Con}

In this paper, incorporating effects of the minimal length, we derived the
deformed Hamilton-Jacobi Equations for both scalars and fermions in curved
spacetime based on the modified fundamental commutation relations. We
investigated the particles' tunneling in the background of a spherically
symmetric black holes. In this spacetime configurations, we showed that the
corrected Hawking temperature is not only determined by the properties of the
black holes, but also dependent on the angular momentum and mass of the
emitted particles. Finally, we studied how a Schwarzschild black hole
evaporates in our model. We found the black hole evaporates to zero mass in
infinite time.

\noindent\textbf{Acknowledgements. }

We would like to acknowledge useful discussions with Y. He, Z. Sun and H. W.
Wu. This work is supported in part by NSFC (Grant No. 11005016, 11175039 and
11375121) and SYSTF (Grant No. 2012JQ0039).

\end{document}